\documentclass{article}
\usepackage{spconf,amsmath,amssymb,graphicx,hyperref}
\usepackage{siunitx}
\usepackage{setspace}
\usepackage{url}

\title{Brain-Informed Speech Separation for Cochlear Implants}

\name{Tom Gajecki$^{(1)}$, Jonas Althoff$^{(1)}$, Waldo Nogueira$^{(1,2)}$}
\address{$^{(1)}$ Hannover Medical School, Department of Otorhinolaryngology; Hearing4all, Hannover, Germany\\
$^{(2)}$ Dept. of Microelectronics \& Electronic Systems, Institut de Neurosci\`encies, UAB, Spain}

\begin{document}
\ninept
\maketitle

\begin{abstract}
We propose a brain-informed speech separation method for cochlear implants (CIs) that uses electroencephalography (EEG)-derived attention cues to guide enhancement toward the attended speaker. An attention-guided network fuses audio mixtures with EEG features through a lightweight fusion layer, producing attended-source electrodograms for CI stimulation while resolving the label-permutation ambiguity of audio-only separators. Robustness to degraded attention cues is improved with a mixed curriculum that varies cue quality during training, yielding stable gains even when EEG-speech correlation is moderate. In multi-talker conditions, the model achieves higher signal-to-interference ratio improvements than an audio-only electrodogram baseline while remaining slightly smaller ($\sim$167k vs.\ $\sim$171k parameters). With 2\,ms algorithmic latency and comparable cost, the approach highlights the promise of coupling auditory and neural cues for cognitively adaptive CI processing.
\end{abstract}

\begin{keywords}
Cochlear Implants, Brain-Computer Interface, EEG, Speech Enhancement, Deep Learning, Curriculum Learning
\end{keywords}

\section{INTRODUCTION}
Cochlear implants (CIs) have revolutionized the treatment of severe to profound hearing loss by providing direct electrical stimulation of the auditory nerve \cite{Lenarz2017}. While modern devices restore reliable speech perception in quiet, CI users still face major challenges in everyday listening environments. A particularly demanding case is the ``cocktail party'' scenario, where multiple talkers overlap and intelligibility often collapses for CI users \cite{cherry1953some}. Conventional sound coding strategies (audio-to-electrode mapping algorithms; e.g., ACE \cite{seligman1995architecture}) primarily map acoustic energy to electrical stimulation patterns and do not explicitly exploit cognitive state or selective attention.

\medskip
\noindent Deep learning has enabled end-to-end CI processing pipelines that produce electrodograms directly from acoustic mixtures \cite{gajecki2022end, gajecki2023deep, electrodenet}. These approaches can suppress background noise and improve robustness, but they remain agnostic to which talker the listener wishes to attend when multiple speech sources overlap. EEG provides a principled source of top-down information: cortical responses selectively track the envelope of the attended talker more strongly than that of competing talkers \cite{mesgarani2012selective, ding2014cortical, osullivan2015singletrial, mirkovic2015decoding}. Leveraging this phenomenon, brain-informed enhancement can, in principl,e steer processing toward the behaviorally relevant talker, which is highly desirable for CIs in realistic multi-talker scenarios \cite{ceolini2020brain, aroudi, ciccarelli2019comparison, vandecappelle2021elife}.

\medskip
\noindent A growing body of work investigates EEG-conditioned speech enhancement or target speaker extraction using real EEG--audio datasets and waveform-domain objectives, including end-to-end brain-driven enhancement \cite{hosseini2022endtoend}, neuro-steered speaker extraction (NeuroHeed) \cite{pan2024neuroheed}, its joint AAD-training extension NeuroHeed+ \cite{pan2024neuroheedplus}, and attention-guided enhancement networks such as BASEN \cite{zhang2023basen} and MSFNet \cite{fan2024msfnet}. More recent approaches explore multimodal representation alignment and multi-scale fusion for brain-assisted extraction (e.g., M3ANet) \cite{fan2025m3anet}. These methods typically output enhanced waveforms and report SI-SDR/SI-SNR-type metrics on EEG--audio corpora. In contrast, our objective is CI-specific: we generate \emph{attended-source electrodograms} suitable for stimulation, building on the lightweight DeepACE family \cite{gajecki2022end, gajecki2023deep}. Because output domains (waveform vs.\ electrodogram), datasets (EEG--audio vs.\ Libri2Mix mixtures), and metrics differ, direct numerical comparisons to waveform-domain neuro-steered models are not strictly apples-to-apples. However, the methodological core is shared: all systems hinge on the reliability of the EEG-derived attention cue. We therefore benchmark against a strong CI electrodogram baseline and focus on robustness as a function of cue quality, which directly targets a main practical bottleneck of EEG-driven systems.

\medskip
\noindent In this work, we introduce an end-to-end brain-informed speech separation system for CIs that integrates attention cues with acoustic mixture processing through a lightweight fusion mechanism \cite{gajecki2023fusion}. The model is designed to remain small and causal (2\,ms latency) to preserve feasibility for hardware-constrained deployment. Since realistic EEG-derived cues are noisy and variable, affected by artifacts, low SNR, and inter-listener variability, training only on idealized cues can overestimate real-world benefits. We address this by explicitly controlling cue degradation during training and evaluation via curriculum learning. The main contributions are:
\begin{enumerate}
\item A compact EEG--acoustic fusion model that produces a \emph{single attended} electrodogram, thus avoiding the label-permutation ambiguity of two-output audio-only separators.
\item A mixed curriculum strategy that exposes the model to both clean and degraded cues throughout training, improving robustness and avoiding overfitting to a narrow cue regime.
\end{enumerate}

\section{METHODS \& MATERIALS}

\subsection{Notation}
Let $x(n)\in\mathbb{R}^{T}$ be a monaural mixture waveform with sample index $n$ and $T$ samples.
We predict an electrodogram $\hat{p}(m,k)$ where $m\in\{1,\dots,M\}$ indexes electrodes ($M=22$) and $k\in\{1,\dots,K\}$ indexes time frames.
Vectors $\hat{p}_{:,k}\in\mathbb{R}^{M}$ denote all electrodes at frame $k$.
The operator $\odot$ denotes element-wise multiplication; $\|\cdot\|$ is the Euclidean norm.
Encoder features have dimension $N\times K$ with $N=64$ channels and $K$ frames.

\subsection{System Architecture}
\begin{figure}[h!]
  \centering
  \includegraphics[width=0.40\textwidth, trim=0.7cm 0.7cm 0.7cm 0.7cm, clip]{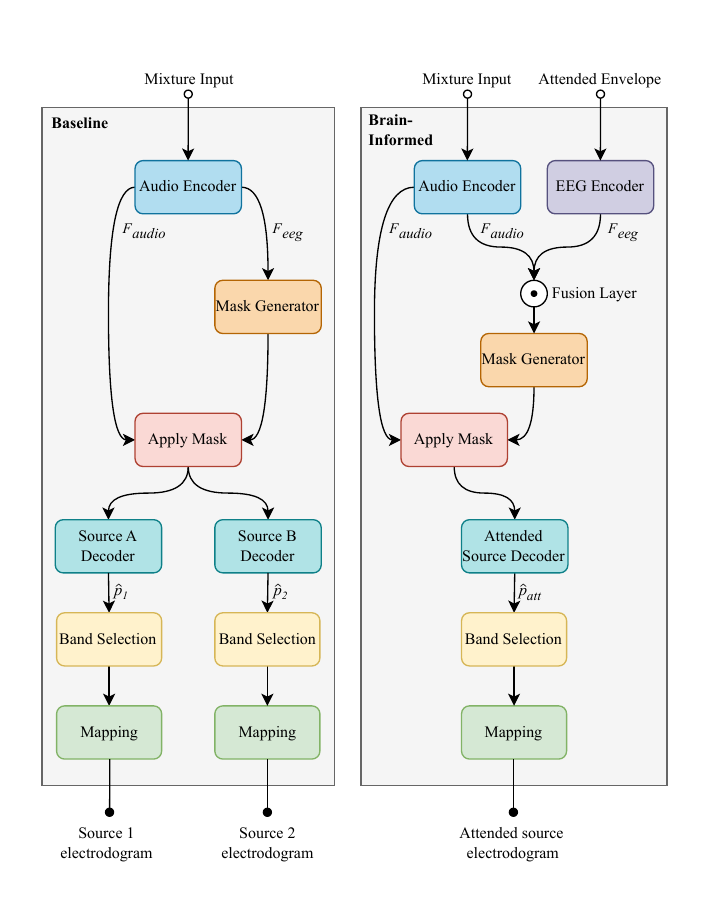}
  \caption{Overview of the baseline speech separation system (left) and brain-informed speech enhancement system (right). The brain-informed system combines audio mixture signals with an EEG-derived attention cue to generate attended-source electrodograms for CI stimulation.}
  \label{system_arch}
\end{figure}

Figure~\ref{system_arch} illustrates the two architectures investigated. The baseline model (left) operates without EEG input and provides two outputs, requiring either an external selector or the listener to choose the desired speaker. In contrast, the brain-informed model (right) uses an EEG-derived attention cue to guide the network toward the attended speaker and produces a single attended electrodogram, thereby resolving output-permutation ambiguity by design. Both models are causal with an algorithmic latency of 2\,ms. The implementation is available as open source at \texttt{Neuro-DeepACE}\footnote{\url{https://github.com/tomgajecki/Neuro-DeepACE}}.

\vspace{-0.6em}
\subsubsection{Baseline Model (audio-only, two outputs)}
The baseline follows the DeepACE architecture for dual-speaker separation \cite{gajecki2022end, gajecki2023deep}. A convolutional encoder $\mathcal{E}$ (kernel $L=32$, stride $L/2$) maps $x(n)$ to features $F\in\mathbb{R}^{N\times K}$. A temporal convolutional network $\mathcal{T}$ (two stacks of dilated blocks; bottleneck $B=64$, hidden $H=128$) produces two masks $\mathcal{M}_1,\mathcal{M}_2\in[0,1]^{N\times K}$, applied as
\[
\tilde{F}_j = F \odot \mathcal{M}_j,\quad j\in\{1,2\}.
\]
A decoder $\mathcal{D}$ maps each $\tilde{F}_j$ to an electrodogram $\hat{p}_j\in[0,1]^{M\times K}$.
As in DeepACE, a post-processing step performs band selection (retaining the $N_{\mathrm{sel}}$ spectral maxima per frame) and maps outputs to stimulation levels using user-specific threshold and comfort levels.
The baseline has $171{,}409$ parameters and is trained with permutation-invariant training (PIT) to address label ambiguity between the two outputs \cite{permutation}.

\vspace{-0.4em}
\subsubsection{Brain-Informed Model (single attended output)}
The brain-informed model augments DeepACE with an EEG-derived attention cue $e(n)$. Encoders of identical architecture but separate parameters yield
\[
F_{\text{audio}}=\mathcal{E}_a(x)\in\mathbb{R}^{N\times K},\qquad
F_{\text{EEG}}=\mathcal{E}_e(e)\in\mathbb{R}^{N\times K}.
\]
A lightweight fusion layer combines both modalities via element-wise multiplication \cite{gajecki2023fusion}:
\[
F_{\text{fused}} = F_{\text{audio}} \odot F_{\text{EEG}}.
\]
The mask generator produces a single mask $\mathcal{M}=\mathcal{T}(F_{\text{fused}})\in[0,1]^{N\times K}$, which is applied to the audio features:
\[
\tilde{F} = F_{\text{audio}} \odot \mathcal{M}.
\]
The decoder yields the attended electrodogram
\[
\hat{p}_{\text{att}} = \mathcal{D}(\tilde{F})\in[0,1]^{M\times K}.
\]
Post-processing (band selection and mapping to stimulation levels) is identical to the baseline. Despite the additional modality, the model remains compact ($167{,}405$ parameters). Because the model outputs a single attended electrodogram, it avoids permutation ambiguity and removes the need for subsequent output selection.

\subsection{Curriculum Learning Strategy}
To improve robustness to degraded EEG--speech association, we investigate three curriculum learning strategies \cite{curriculum}. Here, curriculum learning means controlling the quality of the attention cue during training by injecting noise into the cue: models are first trained on cleaner cues and are then exposed to increasingly degraded cues. This progression is intended to stabilize optimization while improving generalization across cue qualities. We compare:

\noindent\textbf{No Curriculum (oracle cue):}
The attention cue remains clean throughout training. This provides an upper-bound reference that isolates the model's behavior under ideal attention information.

\noindent\textbf{Plain Curriculum:}
Gaussian noise is progressively introduced during training, starting at epoch $n_{\text{start}}=10$ with $\sigma_{\text{init}}=0.05$.
The standard deviation increases by $\Delta\sigma=0.05$ every $T_{\text{step}}=5$ epochs, up to $\sigma_{\text{final}}=0.6$:
\[
\sigma(n) = \min\left(\sigma_{\text{init}} + \left\lfloor \frac{n - n_{\text{start}}}{T_{\text{step}}} \right\rfloor \cdot \Delta\sigma, \, \sigma_{\text{final}}\right).
\]
At epoch $n$, we corrupt the clean attention cue by additive i.i.d.\ Gaussian noise, i.e., $e_{\text{noisy}} = e_{\text{clean}} + \eta$ with $\eta \sim \mathcal{N}(0,\sigma(n)^2)$.

\noindent\textbf{Mixed Curriculum:}
To expose the model to both clean and degraded cues throughout training, we sample the noise standard deviation $\sigma_{\text{mix}}$ at each epoch $n$ according to:
\[
\sigma_{\text{mix}} =
\begin{cases}
0 & \text{with probability } 0.30, \\
\sigma(n) & \text{with probability } 0.65, \\
\mathcal{U}(0,\sigma(n)) & \text{with probability } 0.05,
\end{cases}
\]
where $\mathcal{U}(0,\sigma(n))$ is uniform over $(0,\sigma(n))$.
We then generate $e_{\text{noisy}} = e_{\text{clean}} + \eta$ with $\eta \sim \mathcal{N}(0,\sigma_{\text{mix}}^2)$.
This ensures consistent exposure to clean cues (30\%), the scheduled noise level (65\%), and a small fraction of intermediate noise levels (5\%), which helps stabilize training across cue qualities.
xamples (5\%), which empirically reduces overfitting to a single cue regime.

\subsection{EEG Envelope Processing (proxy attention cue)}
We compute an EEG-like attention cue from the target speech $s(n)$ by rectification and block averaging.
Let $D=\left\lfloor \frac{f_s}{f_{\mathrm{EEG}}} \right\rfloor$ with $f_{\mathrm{EEG}}=64\,\mathrm{Hz}$.
Using non-overlapping windows of length $D$, the downsampled envelope is
\[
e_{\mathrm{d}}(k') \;=\; \frac{1}{D}\sum_{i=0}^{D-1} \bigl|s(k'D+i)\bigr|,
\]
where $k'$ indexes the downsampled frames. We then linearly upsample $e_{\mathrm{d}}(k')$ to the audio sampling rate $f_s$ to obtain $e(n)$ aligned with $s(n)$.

\medskip
\noindent\textbf{Interpretation and limitation.}
The cue above is a \emph{proxy} for an EEG-derived attention envelope rather than a physiologically faithful EEG simulation. In practical neuro-steered pipelines, multi-channel EEG is preprocessed, and an auditory-attention-decoding (AAD) model estimates an attention cue over a time window, which then guides enhancement. Modeling this full chain requires real EEG and specific AAD design choices, which are beyond the scope of this study. Instead, we isolate the downstream question: \emph{given an attention cue of varying reliability, how robustly can the electrodogram generator exploit it?} We therefore control cue degradation via curriculum learning and report performance as a function of cue correlation. Future work will replace the proxy with AAD-derived cues from real EEG and evaluate end-to-end performance under realistic neural noise and artifacts.

\medskip
\noindent\textbf{Attention strength as a function of noise.}
Let $\tilde e = e - \mu(e)$ denote the zero-mean cue and $e_{\mathrm{noisy}}=\tilde e + \eta$ with $\eta\sim\mathcal{N}(0,\sigma_{\mathrm{noise}}^{2})$ independent.
Then the correlation between $e_{\mathrm{noisy}}$ and $\tilde e$ is
\[
\rho \;=\; \mathrm{Corr}(e_{\mathrm{noisy}},\tilde e)
\;=\; \frac{1}{\sqrt{1+\sigma_{\mathrm{noise}}^2/\sigma_{\mathrm{signal}}^2}},
\sigma_{\mathrm{signal}}^2 = \mathrm{Var}(\tilde e).
\]
We use $\rho$ as a scalar summary of cue reliability and present results over a broad range of $\rho$ to emulate the variability observed in practical EEG-driven systems.

\vspace{-0.3em}
\subsection{Loss and Training}
\vspace{-0.2em}
\noindent\textbf{Baseline (two outputs):}
To resolve label ambiguity between two outputs and two targets, we train with permutation-invariant MSE (PIT--MSE):
\[
\mathcal{L}_{\text{PIT}}
=\min_{\pi\in S_2}\frac{1}{K}\sum_{k=1}^{K}
\left(\|\hat p^{(1)}_{:,k}-p^{\mathrm{tar},\pi(1)}_{:,k}\|^2
+\|\hat p^{(2)}_{:,k}-p^{\mathrm{tar},\pi(2)}_{:,k}\|^2\right),
\]
where $S_2$ contains the two permutations and $\pi$ assigns outputs to targets per utterance.

\noindent\textbf{Brain-informed (single attended output):}
Here there is no permutation ambiguity; we minimize standard MSE:
\[
\mathcal{L}_{\text{MSE}}
=\frac{1}{K}\sum_{k=1}^{K}\|p^{\mathrm{tar}}_{:,k}-\hat p_{:,k}\|^2.
\]

\noindent\textbf{Training.}
During training, only the attention cue of the brain-informed model is corrupted according to the selected curriculum (No / Plain / Mixed). The audio-only baseline uses clean audio inputs throughout. Both models are trained with Adam ($10^{-3}$), early stopping (patience 10), and gradient clipping (max-norm 5.0) for up to 100 epochs.

\subsection{Evaluation Metrics}
We quantify performance in the electrodogram domain using:

\[
\mathrm{SIRi}
= 10\log_{10}\frac{\sum_{k=1}^{K}\big\|p^{\mathrm{mix}}_{:,k} - p^{\mathrm{tar}}_{:,k}\big\|^{2}}
{\sum_{k=1}^{K}\big\|\,\hat{p}_{:,k} - p^{\mathrm{tar}}_{:,k}\big\|^{2}},
\]
and electrode-wise linear cross-correlation coefficients (LCCs):
\[
\mathrm{LCC}_m
= \frac{\mathrm{Cov}\!\big(\hat{p}_{m,1:K},\,p^{\mathrm{tar}}_{m,1:K}\big)}
       {\sigma_{\hat{p}_{m,1:K}}\,\sigma_{p^{\mathrm{tar}}_{m,1:K}}}.
\]
Here, $\mathrm{Cov}(\cdot,\cdot)$ is covariance and $\sigma_x$ the standard deviation of $x$.
$p^{\mathrm{mix}}_{:,k}$, $p^{\mathrm{tar}}_{:,k}$, and $\hat{p}_{:,k}$ are the mixture, target, and predicted electrodograms at frame $k$, respectively.
The signal-to-interference ratio improvement (SIRi) measures error reduction relative to the mixture in the stimulation-feature domain, and LCCs provide an interpretable electrode-wise measure of temporal pattern preservation, which is relevant because CI perception depends on temporal envelope cues across channels.

\subsection{Dataset}
We train on two-speaker mixtures from Libri2Mix \cite{cosentino2020librimix} (train-100, 16 kHz). The training set comprises 58 hours from 251 speakers (13.9k mixtures), and the validation/test sets each contain 11 hours from 40 speakers (3k mixtures per split). Mixtures are created with input SIRs uniformly sampled between 0 and 10 dB.

\section{RESULTS}
Results are reported for the attended target; trends were consistent across the evaluated mixtures.

\medskip
\noindent\textbf{Note on comparisons (expanded).}
Our experiments compare (i) a strong audio-only CI electrodogram baseline against (ii) the proposed brain-informed model, and we perform controlled ablations over cue reliability and curriculum strategy. We do not claim direct numerical comparability with waveform-domain neuro-steered models trained/evaluated on real EEG--audio corpora. Instead, the goal is to quantify how much an attention cue can help \emph{CI electrodogram generation} and how robustly the model behaves when the cue becomes unreliable. This complements prior waveform-domain studies by focusing on a CI-relevant output representation and on robustness analysis in terms of cue correlation.

\subsection{Overall Performance Comparison}
\begin{figure}[h!]
\centering
\includegraphics[width = 0.50\textwidth]{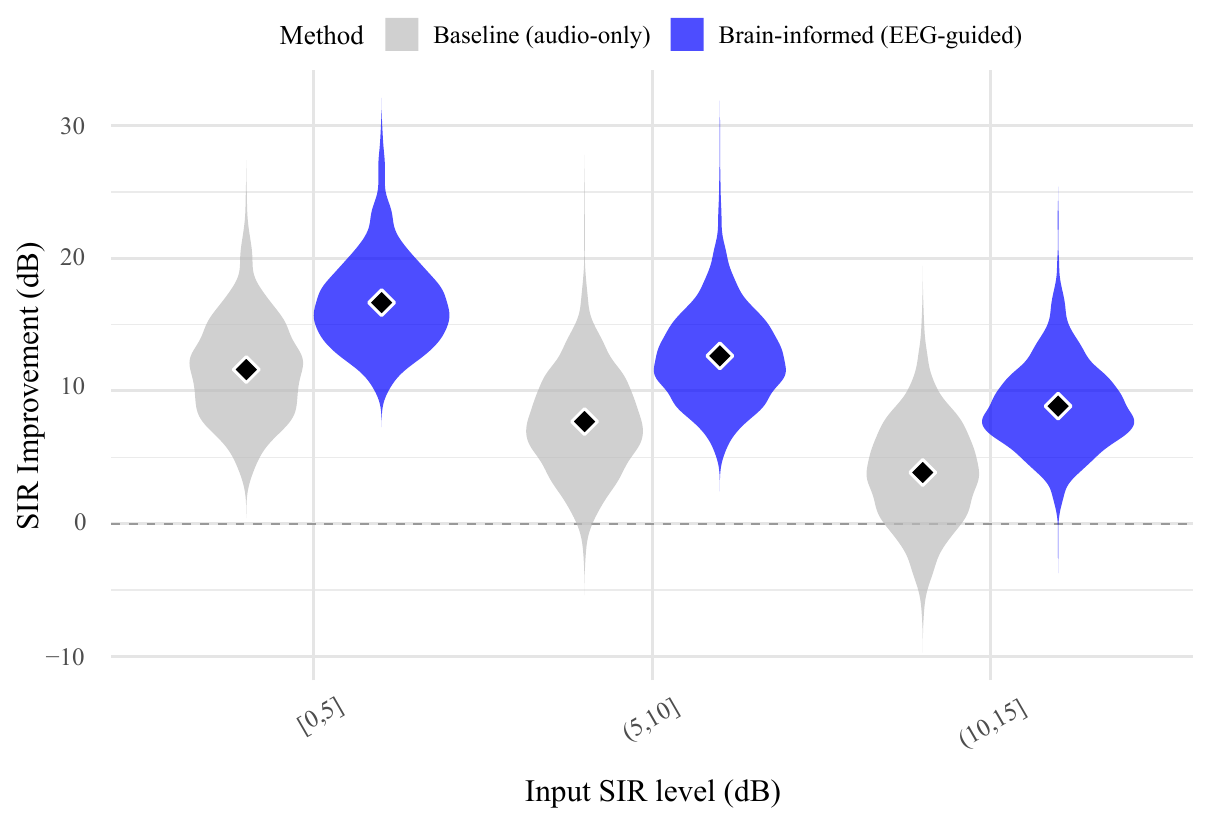}
\caption{SIR improvement of baseline vs.\ brain-informed model across input SIR levels. We assume oracle proxy attention cues for the brain-informed model.}
\label{snr_comparison}
\end{figure}

Figure~\ref{snr_comparison} reports SIR improvements of the baseline and brain-informed system across input SIR levels. With oracle cues (No Curriculum), the brain-informed system consistently outperforms the audio-only baseline at all input SIRs. This indicates that, when attention information is reliable, the fusion mechanism can bias the mask estimation toward the attended source and reduce interference in the predicted electrodograms. The approximately uniform separation across input SIR levels suggests that the benefit is not restricted to a narrow operating point but persists over the tested mixture difficulty range.

\subsection{Cue Correlation Analysis}
\begin{figure}[h!]
\centering
\includegraphics[width = 0.50\textwidth]{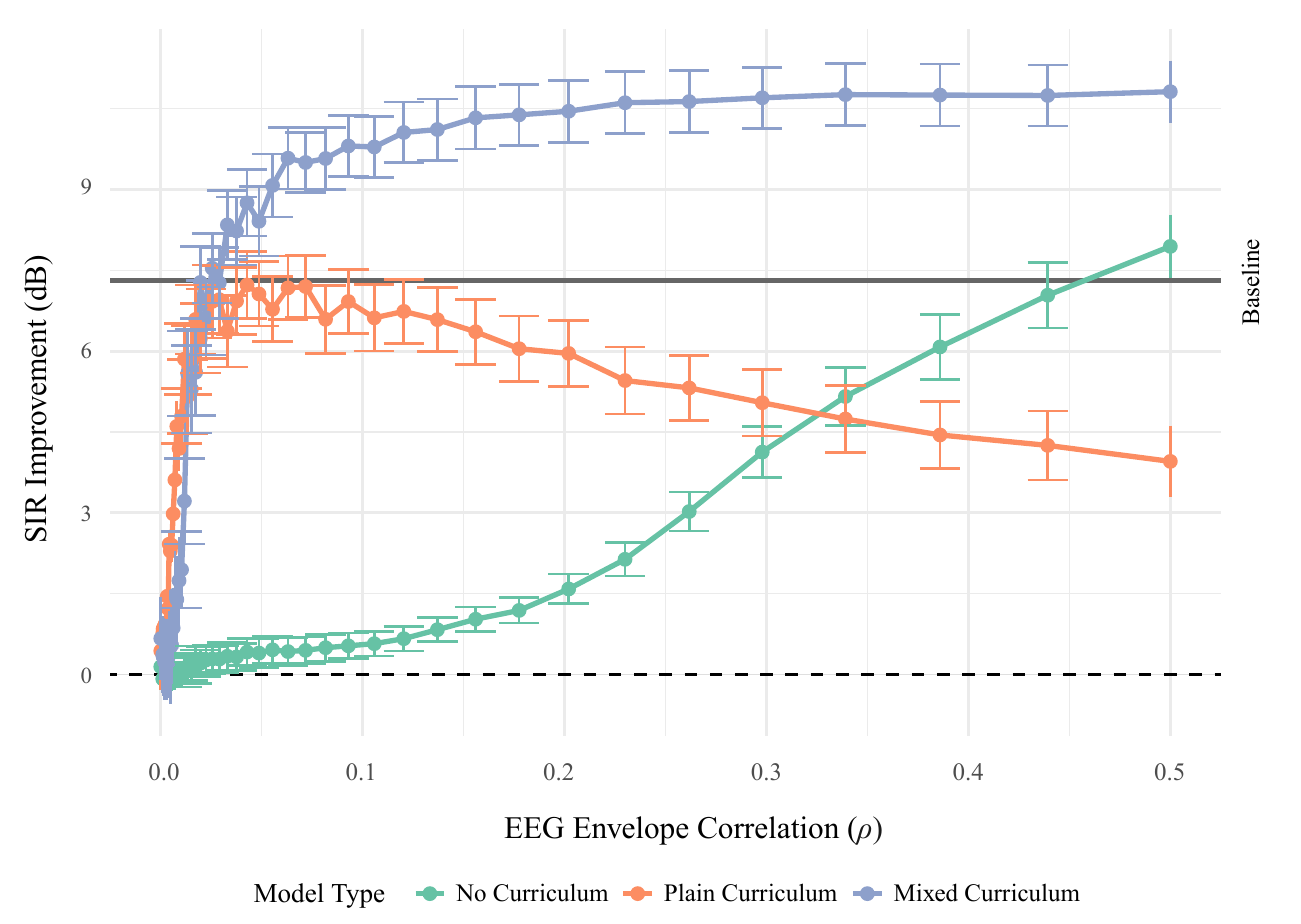}
\caption{SIR improvement as a function of cue correlation for different curriculum strategies. Mixed curriculum shows consistent performance across correlation levels. Error bars indicate standard error.}
\label{eeg_correlation}
\end{figure}

Figure~\ref{eeg_correlation} shows SIR improvement versus cue correlation ($\rho$) under three training strategies (No, Plain, Mixed). SIR generally rises with higher $\rho$, indicating that the model uses cue information to steer enhancement. Importantly, the mapping from $\rho$ to enhancement quality is not fixed; it depends on how the model was exposed to cue degradations during training. This highlights that robustness is not solely determined by architecture but also by the training curriculum.

With the plain curriculum, performance rises with $\rho$ at first but drops at high $\rho$, indicating reduced generalization to clean cues after late-stage training on strongly degraded cues. The mixed curriculum mitigates this by retaining a fraction of clean cues throughout training, yielding more stable gains across $\rho$. Without curriculum, performance increases almost linearly with $\rho$, showing strong dependence on cue quality.

\subsection{Linear Cross-Correlation Analysis}
\begin{figure}[h!]
\centering
\includegraphics[width = 0.50\textwidth]{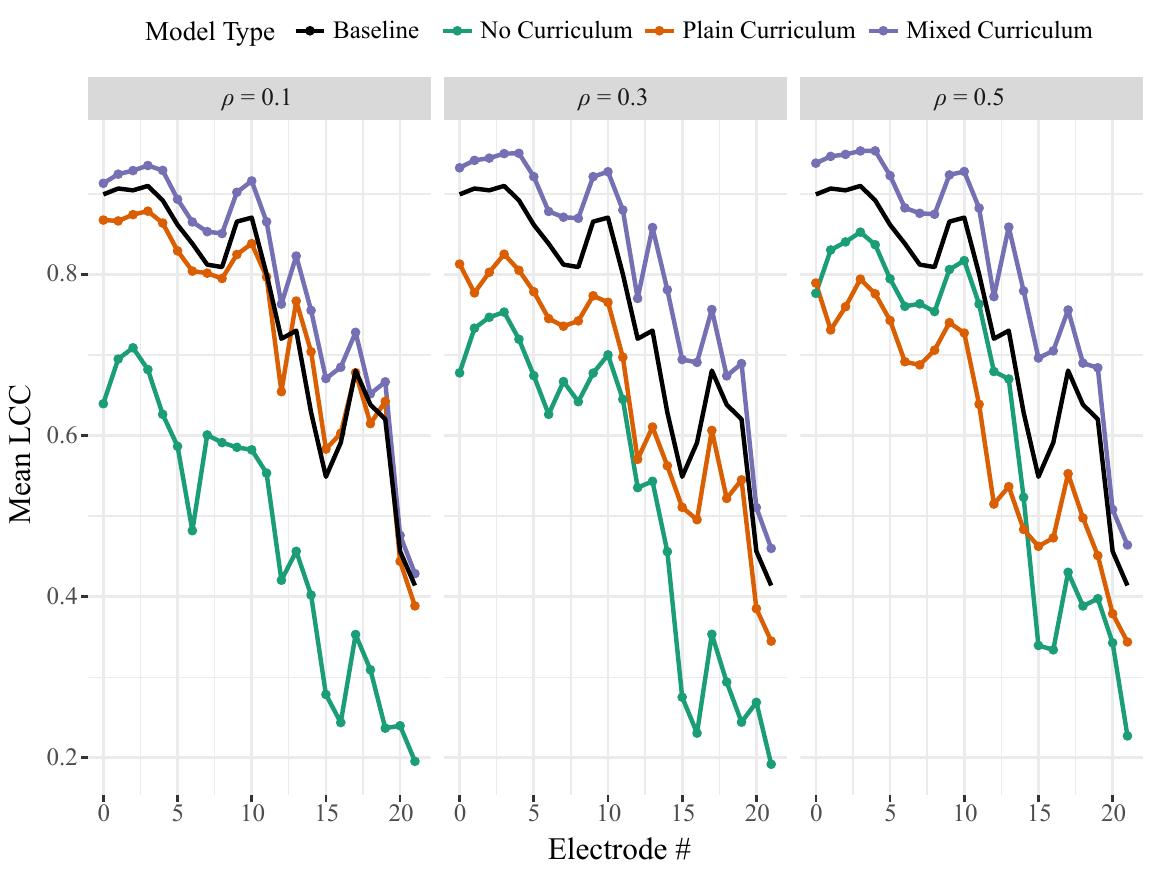}
\caption{LCCs per electrode for different curriculum strategies and the baseline. Higher LCCs indicate stronger preservation of attended-speech temporal patterns, relevant for intelligibility in CI stimulation.}
\label{curriculum_analysis}
\end{figure}

Figure~\ref{curriculum_analysis} shows the impact of curriculum strategy on LCCs across electrodes. Consistent with the correlation analysis, the mixed curriculum achieves the highest LCCs across cue qualities, indicating improved preservation of attended-speech temporal structure in the electrodogram domain. Whereas SIRi reflects aggregate error reduction, LCCs provide electrode-wise insight: higher LCCs suggest that the temporal modulation patterns delivered to each channel more closely match the target, which is relevant for CI speech perception given the importance of envelope cues.

\medskip
\noindent\textbf{Practical interpretation.}
Taken together, the results suggest that (i) reliable attention cues can improve CI electrodogram separation beyond an audio-only baseline, and (ii) training strategy is crucial for robustness when cue reliability varies. In practical EEG-driven deployments, cue reliability is expected to fluctuate over time due to non-stationary artifacts and attention dynamics. The mixed curriculum is therefore attractive as a training principle because it explicitly targets such variability rather than optimizing for a single assumed cue regime.

\vspace{-0.8em}
\section{CONCLUSIONS}
We introduced a brain-informed CI speech-separation model that outperforms an audio-only electrodogram baseline in competing-talker conditions while remaining compact ($\sim$167k vs.\ $\sim$171k parameters) and low latency (2\,ms). A lightweight fusion layer integrates acoustic features with an attention cue to produce a single attended electrodogram, avoiding permutation ambiguity. A mixed curriculum that varies cue quality improves robustness compared with a strictly increasing-noise schedule.

\medskip
\noindent\textbf{Limitations and next steps.}
Our study uses a proxy attention cue and does not include an explicit EEG$\rightarrow$cues/AAD front-end. Future work will validate with real EEG (including CI users), integrate online AAD, and evaluate end-to-end performance under realistic neural noise/artifacts and more complex scenes (e.g., reverberation, additional talkers).

\vfill\pagebreak
\bibliographystyle{IEEEbib}
\bibliography{refs}

@article{cherry1953some,
  title     = {Some Experiments on the Recognition of Speech, with One and with Two Ears},
  author    = {Cherry, Edward Collin},
  journal   = {The Journal of the Acoustical Society of America},
  volume    = {25},
  number    = {5},
  pages     = {975--979},
  year      = {1953},
  publisher = {Acoustical Society of America}
}

@inproceedings{gajecki2022end,
  title     = {An End-to-End Deep Learning Speech Coding and Denoising Strategy for Cochlear Implants},
  author    = {Gajecki, Tom and Nogueira, Waldo},
  booktitle = {ICASSP 2022 IEEE International Conference on Acoustics, Speech and Signal Processing (ICASSP)},
  pages     = {3109--3113},
  year      = {2022},
  organization = {IEEE}
}

@article{seligman1995architecture,
  title     = {Architecture of the Spectra 22 Speech Processor},
  author={Seligman, Peter and McDermott, Hugh},
  journal   = {Annals of Otology, Rhinology \& Laryngology},
  volume    = {104},
  number    = {Suppl 166},
  pages     = {139--141},
  year      = {1995}
}

@article{gajecki2023deep,
  title     = {A Deep Denoising Sound Coding Strategy for Cochlear Implants},
  author    = {Gajecki, Tom and Zhang, Yichi and Nogueira, Waldo},
  journal   = {IEEE Transactions on Biomedical Engineering},
  volume    = {70},
  number    = {9},
  pages     = {2700--2709},
  year      = {2023},
  publisher = {IEEE}
}

@article{Lenarz2017,
  title   = {Cochlear Implant -- State of the Art},
  author  = {Lenarz, Thomas},
  journal = {Laryngorhinootologie},
  volume  = {96},
  number  = {1},
  pages   = {123--151},
  year    = {2017}
}

@article{electrodenet,
  author  = {Huang, Enoch Hsin-Ho and Chao, Rong and Tsao, Yu and Wu, Chao-Min},
  journal = {IEEE Transactions on Cognitive and Developmental Systems},
  title   = {ElectrodeNet --- A Deep-Learning-Based Sound Coding Strategy for Cochlear Implants},
  year    = {2024},
  volume  = {16},
  number  = {1},
  pages   = {346--357},
  doi     = {10.1109/TCDS.2023.3275587}
}

@inproceedings{curriculum,
  author    = {Braun, Stefan and Neil, Daniel and Liu, Shih-Chii},
  booktitle = {2017 25th European Signal Processing Conference (EUSIPCO)},
  title     = {A Curriculum Learning Method for Improved Noise Robustness in Automatic Speech Recognition},
  year      = {2017},
  pages     = {548--552},
  doi       = {10.23919/EUSIPCO.2017.8081267}
}

@article{ceolini2020brain,
  title     = {Brain-Informed Speech Separation ({BISS}) for Enhancement of Target Speaker in Multitalker Speech Perception},
  author    = {Ceolini, Enea and Hjortkj{\ae}r, Jens and Wong, Daniel D. E. and O’Sullivan, James and Raghavan, Vinay S. and Herrero, Jose and Mehta, Ashesh D. and Liu, Shih-Chii and Mesgarani, Nima},
  journal   = {NeuroImage},
  volume    = {223},
  pages     = {117282},
  year      = {2020},
  publisher = {Elsevier}
}

@article{mesgarani2012selective,
  title     = {Selective Cortical Representation of Attended Speaker in Multi-Talker Speech Perception},
  author    = {Mesgarani, Nima and Chang, Edward F.},
  journal   = {Nature},
  volume    = {485},
  number    = {7397},
  pages     = {233--236},
  year      = {2012}
}

@article{ding2014cortical,
  title     = {Cortical Entrainment to Continuous Speech: Functional Roles and Interpretations},
  author    = {Ding, Nai and Simon, Jonathan Z.},
  journal   = {Frontiers in Human Neuroscience},
  volume    = {8},
  pages     = {311},
  year      = {2014}
}

@article{gajecki2023fusion,
  title     = {Deep Latent Fusion Layers for Binaural Speech Enhancement},
  author    = {Gajecki, Tom and Nogueira, Waldo},
  journal   = {IEEE/ACM Transactions on Audio, Speech, and Language Processing},
  volume    = {31},
  pages     = {3127--3138},
  year      = {2023},
  publisher = {IEEE}
}

@misc{cosentino2020librimix,
  title         = {LibriMix: An Open-Source Dataset for Generalizable Speech Separation},
  author        = {Cosentino, Joris and Pariente, Manuel and Cornell, Samuele and Deleforge, Antoine and Vincent, Emmanuel},
  year          = {2020},
  eprint        = {2005.11262},
  archivePrefix = {arXiv},
  primaryClass  = {eess.AS}
}

@article{osullivan2015singletrial,
  title     = {Attentional Selection in a Cocktail Party Environment Can Be Decoded from Single-Trial {EEG}},
  author    = {O'Sullivan, James A. and Power, Alan J. and Mesgarani, Nima and Rajaram, Siddharth and Foxe, John J. and Shinn-Cunningham, Barbara G. and Slaney, Malcolm and Shamma, Shihab A. and Lalor, Edmund C.},
  journal   = {Cerebral Cortex},
  volume    = {25},
  number    = {7},
  pages     = {1697--1706},
  year      = {2015},
  publisher = {Oxford University Press}
}

@article{mirkovic2015decoding,
  title     = {Decoding the Attended Speech Stream with Multi-Channel {EEG}: Implications for Online, Daily-Life Applications},
  author    = {Mirkovic, Bojana and Debener, Stefan and Jaeger, Manuela and De Vos, Maarten},
  journal   = {Journal of Neural Engineering},
  volume    = {12},
  number    = {4},
  pages     = {046007},
  year      = {2015},
  publisher = {IOP Publishing}
}

@article{aroudi,
  author  = {Aroudi, Ali and Doclo, Simon},
  journal = {IEEE/ACM Transactions on Audio, Speech, and Language Processing},
  title   = {Cognitive-Driven Binaural Beamforming Using {EEG}-Based Auditory Attention Decoding},
  year    = {2020},
  volume  = {28},
  pages   = {862--875}
}

@article{ciccarelli2019comparison,
  title     = {Comparison of Two-Talker Attention Decoding from {EEG} with Nonlinear Neural Networks and Linear Methods},
  author    = {Ciccarelli, Gregory and Nolan, Michael and Perricone, Joseph and Calamia, Paul T. and Haro, Stephanie and O’sullivan, James and Mesgarani, Nima and Quatieri, Thomas F. and Smalt, Christopher J.},
  journal   = {Scientific Reports},
  volume    = {9},
  number    = {1},
  pages     = {11538},
  year      = {2019},
  publisher = {Nature Publishing Group UK London}
}

@inproceedings{permutation,
  author    = {Yu, Dong and Kolb{\ae}k, Morten and Tan, Zheng-Hua and Jensen, Jesper},
  booktitle = {2017 IEEE International Conference on Acoustics, Speech and Signal Processing (ICASSP)},
  title     = {Permutation Invariant Training of Deep Models for Speaker-Independent Multi-Talker Speech Separation},
  year      = {2017},
  pages     = {241--245},
  doi       = {10.1109/ICASSP.2017.7952154}
}

@article{vandecappelle2021elife,
  title     = {{EEG}-Based Detection of the Locus of Auditory Attention with Convolutional Neural Networks},
  author    = {Vandecappelle, Servaas and Deckers, Lucas and Das, Neetha and Ansari, Amir Hossein and Bertrand, Alexander and Francart, Tom},
  journal   = {eLife},
  volume    = {10},
  pages     = {e56481},
  year      = {2021},
  publisher = {eLife Sciences Publications, Ltd}
}

@article{hosseini2022endtoend,
  author  = {Maryam Hosseini and Luca Celotti and {\'E}ric Plourde},
  title   = {End-to-End Brain-Driven Speech Enhancement in Multi-Talker Conditions},
  journal = {IEEE/ACM Transactions on Audio, Speech, and Language Processing},
  volume  = {30},
  pages   = {1718--1733},
  year    = {2022},
  doi     = {10.1109/TASLP.2022.3169629}
}

@article{pan2024neuroheed,
  author  = {Zexu Pan and Marvin Borsdorf and Siqi Cai and Tanja Schultz and Haizhou Li},
  title   = {NeuroHeed: Neuro-Steered Speaker Extraction Using EEG Signals},
  journal = {IEEE/ACM Transactions on Audio, Speech, and Language Processing},
  volume  = {32},
  pages   = {4456--4470},
  year    = {2024},
  doi     = {10.1109/TASLP.2024.3463498}
}

@inproceedings{pan2024neuroheedplus,
  author    = {Zexu Pan and Gordon Wichern and Fran{\c{c}}ois G. Germain and Sameer Khurana and Jonathan Le Roux},
  title     = {NeuroHeed+: Improving Neuro-Steered Speaker Extraction with Joint Auditory Attention Detection},
  booktitle = {Proc. IEEE International Conference on Acoustics, Speech and Signal Processing (ICASSP)},
  year      = {2024}
}

@inproceedings{zhang2023basen,
  author    = {Jie Zhang and Qing-Tian Xu and Qiu-Shi Zhu and Zhen-Hua Ling},
  title     = {{BASEN}: Time-Domain Brain-Assisted Speech Enhancement Network with Convolutional Cross Attention in Multi-talker Conditions},
  booktitle = {Proc. Interspeech},
  pages     = {3117--3121},
  year      = {2023},
  doi       = {10.21437/Interspeech.2023-673}
}

@inproceedings{fan2024msfnet,
  author    = {Cunhang Fan and Jingjing Zhang and Hongyu Zhang and Wang Xiang and Jianhua Tao and Xinhui Li and Jiangyan Yi and Dianbo Sui and Zhao Lv},
  title     = {{MSFNet}: Multi-Scale Fusion Network for Brain-Controlled Speaker Extraction},
  booktitle = {Proceedings of the 32nd ACM International Conference on Multimedia (ACM MM)},
  pages     = {1652--1661},
  year      = {2024}
}

@inproceedings{fan2025m3anet,
  author    = {Cunhang Fan and Ying Chen and Jian Zhou and Zexu Pan and Jingjing Zhang and Youdian Gao and Xiaoke Yang and Zhengqi Wen and Zhao Lv},
  title     = {{M3ANet}: Multi-scale and Multi-Modal Alignment Network for Brain-Assisted Target Speaker Extraction},
  booktitle = {Proceedings of the Thirty-Fourth International Joint Conference on Artificial Intelligence (IJCAI-25)},
  year      = {2025}
}

\end{document}